\documentclass [aps,prl,twocolumn,showpacs,nofootinbib,superscriptaddress]{revtex4}
\usepackage{epsfig}
\usepackage{graphicx}
\usepackage{amsmath}
\usepackage{mathrsfs}
\usepackage{bm}

\begin{document}

\preprint{DESY~13--057, UWTHPH--2013--6\hspace{9.8cm}ISSN 0418-9833}
\preprint{March 2013\hspace{16.2cm}}

\boldmath
\title{Next-to-leading-order nonrelativisitic QCD disfavors interpretation of
$X(3872)$ as $\chi_{c1}(2P)$}
\unboldmath


\author{Mathias Butenschoen}
\affiliation{Universit\"at Wien, Fakult\"at f\"ur Physik,
Boltzmanngasse 5, 1090 Wien, Austria}

\author{Zhi-Guo He}
\affiliation{II. Institut f\"ur Theoretische Physik, Universit\"at Hamburg,
Luruper Chaussee 149, 22761 Hamburg, Germany}
\author{Bernd A. Kniehl}
\affiliation{II. Institut f\"ur Theoretische Physik, Universit\"at Hamburg,
Luruper Chaussee 149, 22761 Hamburg, Germany}

\date{\today}

\begin{abstract}
We study $\chi_{c1}(2P)$ inclusive hadroproduction at next-to-leading order
(NLO), both in $\alpha_s$ and $v^2$, within the factorization formalism of
nonrelativistic quantum chromodynamics (NRQCD), including the color-singlet
$^3\!P_1^{[1]}$ and color-octet $^3\!S_1^{[8]}$ $c\bar c$ Fock states as well
as the mixing of the latter with the $^3\!D_1^{[8]}$ state.
Assuming the recently discovered $X(3872)$ hadron to be the $J^{PC}=1^{++}$
charmonium state $\chi_{c1}(2P)$, we perform a fit to the cross sections
measured by the CDF, CMS, and LHCb Collaborations.
We either obtain an unacceptably high value of $\chi^2$, a value of
$|R_{2P}^\prime(0)|$ incompatible with well-established potential models, or an
intolerable violation of the NRQCD velocity rules.
We thus conclude that NLO NRQCD is inconsistent with the hypothesis
$X(3872)\equiv\chi_{c1}(2P)$.
\end{abstract}


\maketitle

During the past decade, a series of charmonium or charmonium-like $X,Y,Z$
states were discovered (for a recent review, see Ref.~\cite{Faccini:2012pj}).
The $X(3872)$ state is one of the most interesting among them.
It was discovered in 2003 by the Belle Collaboration at KEKB in $B$ meson
decays \cite{Choi:2003ue}, and confirmed shortly afterwards by the BaBar
Collaboration at SLAC PEP-II \cite{Aubert:2004ns}.
It was also observed by the CDF \cite{Acosta:2003zx} and D0
\cite{Abazov:2004kp} Collaborations in $p\bar{p}$ collisions at the Tevatron
Fermilab.
Ever since its discovery, many theoretical group have tried to interpret its
nature, which has remained mysterious to date, and it is an urgent task of
great importance and broad interest to solve this notorious puzzle of hadron
spectroscopy.
Typical options include conventional charmonia \cite{Barnes:2003vb},
$D^{\ast0}\overline{D}^0/D^0\overline{D}^{\ast0}$ molecules \cite{Close:2003sg},
and tetraquarks \cite{Maiani:2004vq}.
However, none of them can provide a convincing description of all the
experimental measurements.
After analyzing the dipion mass spectrum in $X(3872) \to J/\psi+\pi^+\pi^-$,
only two options for its $J^{PC}$ property are left, either $1^{++}$ or
$2^{-+}$ \cite{Abulencia:2005zc}.\footnote{%
Very recently,  the LHCb Collaboration \cite{Aaij:2013zoa} established the
assignment $J^{PC}=1^{++}$, which, however, still lacks independent
confirmation.}
In $p\bar{p}$ and $pp$ collisions, most of the $X(3872)$ mesons are produced
promptly rather than through decays of $b$ hadrons
\cite{Bauer:2004bc,Chatrchyan:2013cld}.
The study of $X(3872)$ prompt production provides complementary information on
its nature.
In Ref.~\cite{Bignamini:2009sk}, the cross section of $X(3872)$ was estimated
under the assumption that it is a loosely-bound
$D^{\ast0}\overline{D}^0/D^0\overline{D}^{\ast0}$ molecule, and the upper bound of
the theoretical calculation was found to be much smaller than the CDF
measurement \cite{Acosta:2003zx,Bauer:2004bc}.
Later, Artoisenet and Braaten \cite{Artoisenet:2009wk} pointed out that the
upper bound of this prediction can be rendered consistent with the Tevatron
data \cite{Acosta:2003zx,Abazov:2004kp,Bauer:2004bc} by properly taking into
account rescattering effects.
They also used the NRQCD factorization approach \cite{Bodwin:1994jh} to
interpret $X(3872)$ prompt production at the Tevatron
\cite{Acosta:2003zx,Abazov:2004kp,Bauer:2004bc} and presented predictions for
the LHC.
However, their predictions significantly exceed the new measurements reported
by the CMS \cite{Chatrchyan:2013cld} and LHCb \cite{Aaij:2011sn}
Collaborations.
In their charmonium interpretation \cite{Artoisenet:2009wk}, $X(3872)$ is
assumed to be a $1^{++}$ state that is dominantly produced via the color-octet
$c\bar{c}$ Fock state $^3\!S_1^{[8]}$, and the short-distance coefficients are
calculated at leading order (LO).
Note that, at first sight, the mass value 3.872~GeV seems too low for a
$\chi_{c1}(2P)$ candidate, but color-screening effects together with
coupled-channel effects may draw its mass down towards 3.872~GeV
\cite{Li:2009zu}.

Recent NRQCD analyses have revealed that NLO corrections play a key role in
explaining the $J/\psi$ \cite{Ma:2010yw} and $\chi_{cJ}(1P)$ \cite{Ma:2010vd}
yields measured at the Tevatron and the LHC.
Under the assumption that $X(3872)$ is the $1^{++}$ charmonium state
$\chi_{c1}(2P)$, it is then natural to ask if its prompt production rates may
be explained upon including NLO corrections.
The main goal of our work is to answer this question.
To this end, we shall first calculate the cross section of inclusive
$\chi_{c1}(2P)$ hadroproduction at NLO in NRQCD and then check if its free
parameters can be adjusted so as to yield a satisfactory description of the
available prompt $X(3872)$ hadroproduction data
\cite{Acosta:2003zx,Bauer:2004bc,Chatrchyan:2013cld,Aaij:2011sn}.

Owing to the factorization theorems of the QCD parton model and NRQCD
\cite{Bodwin:1994jh}, the inclusive $\chi_{c1}(2P)$ hadroproduction cross
section is evaluated from
\begin{eqnarray}\label{xs}
&&d\sigma(AB\to \chi_{c1}(2P)+X)
=\sum_{i,j,n} \int dxdy\, f_{i/A}(x)f_{j/B}(y)
\nonumber\\
&&{}\times\langle{\cal O}^{\chi_{c1}(2P)}[n]\rangle
d\sigma(i j\to c\overline{c}[n]+X),
\end{eqnarray}
where $f_{i/A}(x)$ are the parton distribution functions (PDFs) of hadron $A$,
$\langle{\cal O}^{\chi_{c1}(2P)}[n]\rangle$ are the long-distance
matrix elements (LDMEs), and
$d\sigma(i j\to c\overline{c}[n]+X)$ are the partonic cross sections.
Working in the fixed-flavor-number scheme, $i$ and $j$ run over the gluon $g$
and the light quarks $q=u,d,s$ and anti-quarks $\overline q$.
The system $X$ always contains one hard parton at LO and is taken to be devoid
of heavy flavors, which may be tagged and vetoed experimentally.
The contribution due to final states in which $X$ comprises an open $c\bar{c}$
pair is found to be suppressed by one order of magnitude \cite{Li:2011yc}.
At LO in the relative velocity $v$ of the bound $c$ and $\bar c$ quarks in the
charmonium rest frame, only the states $n={}^3\!P_1^{[1]},{}^3\!S_1^{[8]}$
contribute \cite{Bodwin:1994jh}.
We evaluate the NLO corrections, which are of relative orders
$\mathcal{O}(\alpha_s)$ and $\mathcal{O}(v^2)$.

In our $\mathcal{O}(\alpha_s)$ calculation, all singularities are canceled
analytically.
The ultraviolet divergences are removed by renormalizing the parameters
$\alpha_s$ and $m_c$ and the wave functions of the external lines.
Specifically, we work in the on-shell scheme, except for $\alpha_s$, which is
treated in the 
$\overline{\mathrm{MS}}$ scheme.
The infrared 
singularities are canceled similarly as described in
Ref.~\cite{Butenschoen:2009zy}.
Notice that the inclusion of the $^3\!S_1^{[8]}$ contribution is indispensable
in order to obtain an IR finite result.
We thus recover the notion that the color-singlet model is not a complete
theory.
By the same token, the dependencies of
$\langle{\cal O}^{\chi_{c1}(2P)}[n]\rangle$ and
$d\sigma(i j\to c\overline{c}[n]+X)$ on the NRQCD factorization scale
$\mu_\Lambda$ only cancel after summation over $n$.
The $\mathcal{O}(v^2)$ corrections involve the additional
$\langle\mathcal{P}^{\chi_{c1}(2P)}(^3\!P_1^{[1]})\rangle$,
$\langle\mathcal{P}^{\chi_{c1}(2P)}(^3\!S_1^{[8]})\rangle$, and
$\langle\mathcal{P}^{\chi_{c1}(2P)}(^3\!S_1^{[8]},^3\!D_1^{[8]})\rangle$
LDMEs of the respective local four-fermion operators of mass dimension eight
\cite{Bodwin:1994jh} and may be evaluated from tree-level diagrams of $c\bar c$
hadroproduction similarly as for hadronic quarkonium decays
\cite{Brambilla:2008zg}.

We now describe the choices of input for our NLO NRQCD calculation. 
We take the charm-quark mass to be $m_c=1.5$~GeV and use the two-loop formula
for $\alpha_s^{(n_f)}$ with $n_f=4$ active quark flavors.
As for the proton PDFs, we adopt the CTEQ6M set \cite{Pumplin:2002vw}, which
comes with asymptotic scale parameter $\Lambda_{\mathrm{QCD}}^{(4)}=326$~MeV.
We choose the $\overline{\mathrm{MS}}$ renormalization, factorization, and
NRQCD scales to be $\mu_r=\mu_f=\xi m_T$ and $\mu_{\Lambda}=\eta m_c$, where
$m_T=\sqrt{p_T^2+4m_c^2}$ is the $\chi_{c1}(2P)$ transverse mass, and
independently vary $\xi$ and $\eta$ by a factor of two up and down about their
default values $\xi=\eta=1$ to estimate the scale uncertainty.
To LO in $v$, we have \cite{Bodwin:1994jh}
\begin{equation}
\langle\mathcal{O}^{\chi_{c1}(2P)}(^3\!P_1^{[1]})\rangle
=(2J+1)\frac{3C_{A}}{2\pi}|R_{2P}^{\prime}(0)|^2,
\label{LDME}
\end{equation}
where $C_A=N_c=3$, $J=1$ is the total angular momentum of the $\chi_{c1}(2P)$
meson and $R_{2P}(r)$ is its radial wave function, which may be calculated
using models for the QCD potential of the charm quark.
Adopting frequently used potential models with different choices of parameters,
$|R_{2P}^{\prime}(0)|^2$ is found to range from 0.076~GeV$^5$ to 0.183~GeV$^5$
\cite{Eichten:1995ch}.
As the default for our fits, we adopt the value
$|R_{2P}^{\prime}(0)|^2=0.102$~GeV$^5$ obtained using the Buchm\"uller-Tye
potential \cite{Buchmuller:1980su}.
To compare theoretical predictions with the experimental data, we also need to
know the branching fraction (BR) of the decay mode
$X(3872)\to J/\psi+\pi^{+}\pi^{-}$ used to identify the $X(3872)$ meson.
It has not been determined yet, but the lower bound $\mathrm{BR}>2.6\%$
has been established at 90\% C.L. \cite{Beringer:1900zz}.
Furthermore, the upper bound $\mathrm{BR}<9.3\%$ was derived at $90\%$ C.L.
using constrains from some other decay channels \cite{Artoisenet:2009wk}.
In our fits, we use $\mathrm{BR}=2.6\%$.

Based on the measurements by the CDF Collaboration
\cite{Acosta:2003zx,Bauer:2004bc}, at center-of-mass energy
$\sqrt{s}=1.96$~TeV, the prompt production cross section of $X(3872)$ mesons
with rapidity $|y|<0.6$ and transverse momentum $p_T>5$~GeV is estimated to be
\cite{Bignamini:2009sk,Artoisenet:2009wk}
\begin{equation}\label{CDF}
\sigma_{\mathrm{CDF}}^{\mathrm{prompt}}(p\bar{p}\to X(3872)+X)\mathrm{BR}
=(3.1\pm0.7)~\mathrm{nb}.
\end{equation}
At LHC, prompt $X(3872)$ production was first measured by the CMS Collaboration
\cite{Chatrchyan:2013cld}, at $\sqrt{s}=7$~TeV, with the result
\begin{equation}\label{CMS}
\sigma_{\mathrm{CMS}}^{\mathrm{prompt}}(pp\to X(3872)+X)\mathrm{BR}
=(1.06\pm0.19)~\mathrm{nb},
\end{equation}
for $|y|<1.2$ and 10~GeV${}< p_T <30$~GeV.
They also presented a $p_T$ distribution \cite{Chatrchyan:2013cld}.
The LHCb Collaboration also measured $X(3872)$ production at $\sqrt{s}=7$~TeV,
but did not discriminate between $b$-hadron and prompt sources
\cite{Aaij:2011sn}.
Averaging the CDF and CMS measurements of the non-prompt fraction,
$(16.1\pm4.9\pm2.0)\%$ \cite{Acosta:2003zx,Bauer:2004bc} and
$(26.3\pm2.3\pm1.6)\%$ \cite{Chatrchyan:2013cld}, respectively, we estimate
the LHCb prompt cross section to be
\begin{equation}\label{LHCb}
\sigma_{\mathrm{LHCb}}^{\mathrm{prompt}}(pp\to X(3872)+X)\mathrm{BR}
=(4.26\pm1.23)~\mathrm{nb},
\end{equation}
for $2.5<y<4.5$ and 5~GeV${}<p_T<20$~GeV.

\begin{table*}
\caption{Results of our $\mathcal{O}(\alpha_s)$ NLO NRQCD fits to the measured
$p_T$ distribution of prompt $X(3872)$ production \cite{Chatrchyan:2013cld} and
the integrated cross section of Eq.~(\ref{CDF}) including or excluding the
result of Eq.~(\ref{LHCb}).
In the one-parameter case,  
$\langle\mathcal{O}^{\chi_{c1}(2P)}(^3\!P_{1}^{[1]})\rangle$ is determined by
the potential model of Ref.~\cite{Buchmuller:1980su}, while it is a fit
parameter in the two-parameter case.
We adopt $\mathrm{BR}=2.6\%$.
}\label{Table}
\small
\begin{tabular}{c|cc|cc}
 & \multicolumn{2}{|c}{One-parameter fit} &
\multicolumn{2}{|c}{Two-parameter fit} \\
 & w/ LHCb data & w/o LHCb data & w/ LHCb data & w/o LHCb data \\
\hline
$\langle\mathcal{O}^{\chi_{c1}(2P)}(^3\!P_{1}^{[1]})\rangle$
$[\mathrm{GeV}^{5}]$ & 0.438 & 0.438 & $0.100^{+0.050}_{-0.050}$ & 
$0.190^{+0.092}_{-0.094}$ \\
$\langle\mathcal{O}^{\chi_{c1}(2P)}(^3\!S_{1}^{[8]})\rangle$
$[\mathrm{GeV}^{3}]$ &
$(3.84^{+0.28}_{-0.24})\times10^{-3}$ & $(4.30^{+0.30}_{-0.26})\times10^{-3}$ &
$(2.95^{+0.54}_{-0.58})\times10^{-3}$ & $(3.36^{+0.56}_{-0.66})\times10^{-3}$\\
$\chi^2/\mathrm{d.o.f.}$ & $79.1/5=15.8$ & $16.7/4=4.18$ & $4.26/4=1.07$ &
$0.63/3=0.21$ \\ 
\end{tabular}
\end{table*}

In the following, we perform a NLO NRQCD test of the hypothesis that the
$X(3872)$ hadron is the $\chi_{c1}(2P)$ charmonium state.
Since $X(3872)$ production via feed-down of heavier charmonia has not been
observed, we assume for the time being prompt production to be approximately
exhausted by direct production.
In fact, charmonia heavier than the $X(3872)$ hadron have sufficient phase
space above the open charm production threshold and preferably decay to pairs
of $D$ mesons so as to evade the kinematic constraint of $c\bar c$ bound state
formation. 
To start with, we neglect the $\mathcal{O}(v^2)$ corrections, which will be
studied in a second step.
We are thus led to fit Eq.~(\ref{xs}), which depends on the parameters
$\langle\mathcal{O}^{\chi_{c1}(2P)}(^3\!P_1^{[1]})\rangle$ and
$\langle\mathcal{O}^{\chi_{c1}(2P)}(^3\!S_1^{[8]})\rangle$, to the experimental
data of prompt $X(3872)$ production
\cite{Acosta:2003zx,Bauer:2004bc,Chatrchyan:2013cld,Aaij:2011sn}.
We consider four options altogether.
On the theoretical side, we either fix
$\langle\mathcal{O}^{\chi_{c1}(2P)}(^3\!P_1^{[1]})\rangle=0.438$~GeV$^5$ by the
potential model of Ref.~\cite{Buchmuller:1980su}, or fit it along with
$\langle\mathcal{O}^{\chi_{c1}(2P)}(^3\!S_1^{[8]})\rangle$.
In the latter case, we actually take
$\langle\mathcal{O}^{\chi_{c1}(2P)}(^3\!P_1^{[1]})\rangle\mathrm{BR}$ and
$\langle\mathcal{O}^{\chi_{c1}(2P)}(^3\!S_1^{[8]})\rangle\mathrm{BR}$ to be fit
parameters.
On the experimental side, we either include the LHCb result \cite{Aaij:2011sn}
of Eq.~(\ref{LHCb}) in the fit along with the CDF result
\cite{Acosta:2003zx,Bauer:2004bc} of Eq.~(\ref{CDF}) and the CMS measurement of
the $p_T$ distribution \cite{Chatrchyan:2013cld}, which includes four data
points, or we exclude it.
In order to avoid double counting, we always exclude the CMS result
\cite{Chatrchyan:2013cld} of Eq.~(\ref{CMS}) from the fit.

\begin{figure*}
\begin{center}
\begin{tabular}{cc}
\includegraphics[scale=0.80]{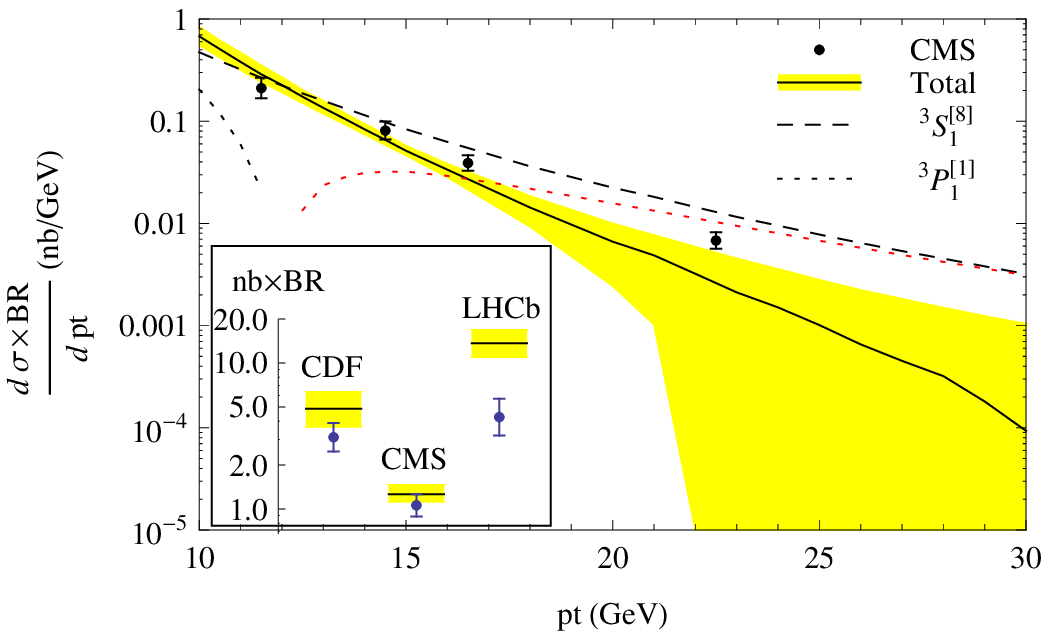}&
\includegraphics[scale=0.80]{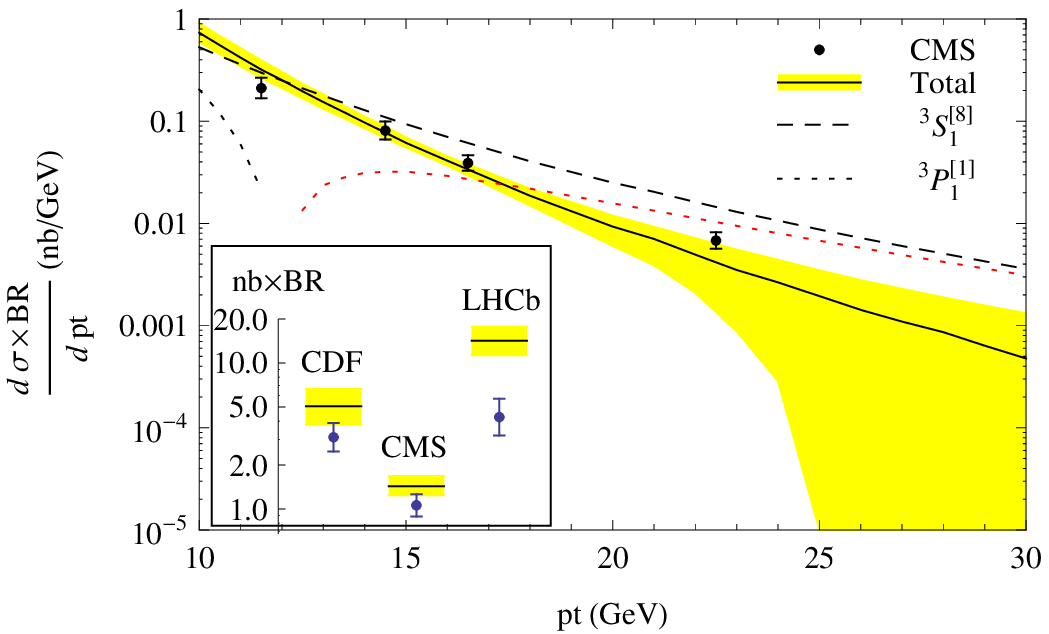}\\
\includegraphics[scale=0.80]{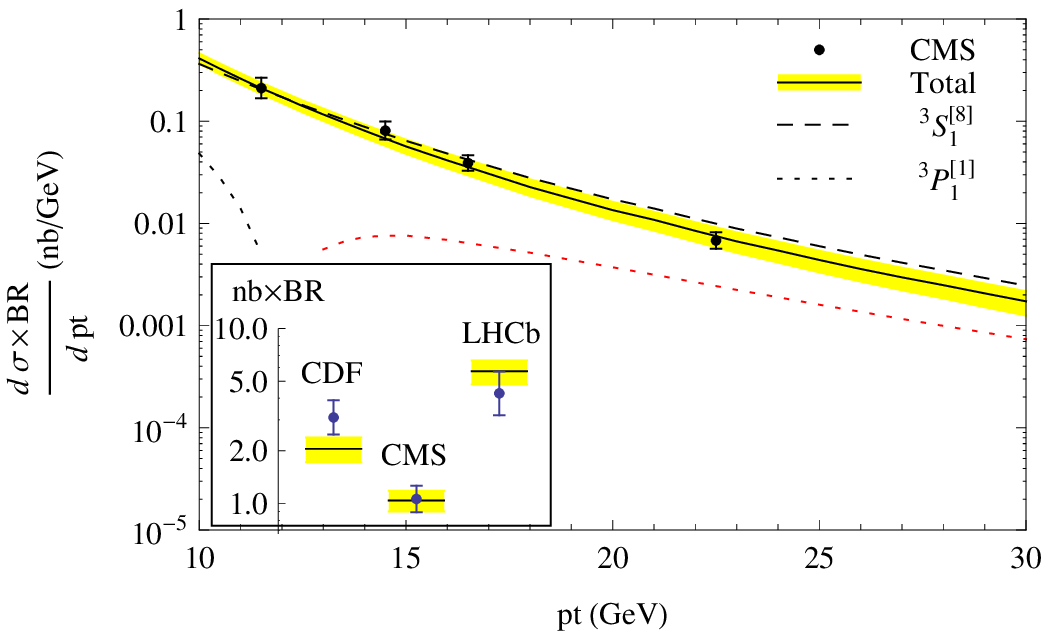}&
\includegraphics[scale=0.80]{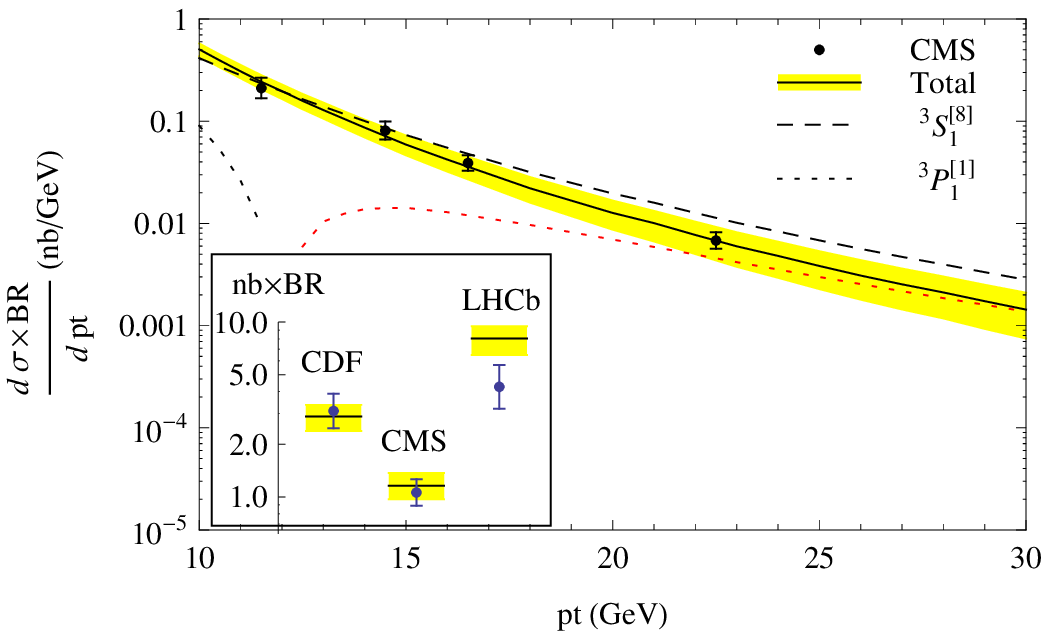}
\end{tabular}
\caption{The prompt $X(3872)$ production cross sections measured by the CDF
\cite{Acosta:2003zx,Bauer:2004bc}, CMS \cite{Chatrchyan:2013cld}, and LHCb
\cite{Aaij:2011sn} Collaborations are compared with NLO NRQCD results based on
one-parameter (upper row) or two-parameter (lower row) fits including (left
column) or excluding (right column) the LHCb data point of Eq.~(\ref{LHCb})
\cite{Aaij:2011sn}.
Dotted, dashed, and solid lines represent the $^3\!P_1^{[1]}$ and
$^3\!S_1^{[8]}$ contributions and their sum, respectively.
Grey/red lines denote negative values, familiar from
Refs.~\cite{Ma:2010yw,Ma:2010vd}.
Shaded/yellow bands indicated the uncertainties in the total results.}
\label{Plot}
\end{center}
\end{figure*}

The results of the four fits are summarized in Table~\ref{Table}, and their
goodness may be conveniently assessed from Fig.~\ref{Plot}.
In the two-parameter case, the quoted values of
$\langle\mathcal{O}^{\chi_{c1}(2P)}(^3\!P_{1}^{[1]})\rangle$ and
$\langle\mathcal{O}^{\chi_{c1}(2P)}(^3\!S_{1}^{[8]})\rangle$ correspond to our
default value $\mathrm{BR}=2.6\%$.
When Eq.~(\ref{LHCb}) is excluded from the fit,
$\sigma^{\mathrm{prompt}}_{\mathrm{LHCb}}\mathrm{BR}$ is a genuine prediction.
The uncertainties are estimated by adding in quadrature the errors of
experimental origin resulting from the fits using our default NLO NRQCD
results and those due the variations of the scale parameters $\xi$ and $\eta$.

The one-parameter fit including the LHCb data point of Eq.~(\ref{LHCb}) has
$\mathrm{d.o.f.}=5$ degrees of freedom and yields $\chi^2=79.1$, so that
$\chi^2/\mathrm{d.o.f.}=15.8$ is intolerably large suggesting that the
experimental data is poorly described by only adjusting
$\langle\mathcal{O}^{\chi_{c1}(2P)}(^3\!S_{1}^{[8]})\rangle$.
This is also evident from the upper left panel in Fig.~\ref{Plot}.
We observe that $\chi^2$ rapidly increases with $|R_{2P}^\prime(0)|^2$ and BR.
For $|R_{2P}^\prime(0)|=0.076$~GeV$^5$ and $\mathrm{BR}=2.6\%$, we obtain the
best value $\chi^{2}=37.5$, which is still unacceptably large, while for
$|R_{2P}^\prime(0)|^2=0.183$~GeV$^5$ and $\mathrm{BR}=9.3\%$, $\chi^2$ is
around 5000.
We also notice that the NLO NRQCD result greatly overshoots the LHCb data point
although it is included in the fit.
Excluding it from the fit mildly increases the central values of the fit
results and their errors, but significantly reduces $\chi^2$, by almost a
factor of five.
The NLO NRQCD prediction for Eq.~(\ref{LHCb}) is then
$(14.21^{+3.41}_{-2.83})$~nb, so that the theory band overshoots the LHCb
result by almost six experimental standard deviations.

The two-parameter fit including the LHCb data point of Eq.~(\ref{LHCb}) works
nicely, yielding $\chi^2/\mathrm{d.o.f.}=1.07$. 
Specifically, we have
$\langle\mathcal{O}^{\chi_{c1}(2P)}(^3\!P_1^{[1]})\rangle\mathrm{BR}=
(2.60^{+1.30}_{-1.30})\times10^{-3}$~GeV$^5$ and
$\langle\mathcal{O}^{\chi_{c1}(2P)}(^3\!S_1^{[8]})\rangle\mathrm{BR}=
(7.67^{+1.40}_{-1.51})\times10^{-5}$~GeV$^3$.
From $\mathrm{BR}>2.6\%$ at 90\% C.L. \cite{Beringer:1900zz}, we thus derive
the 90\% C.L. upper bound
$|R_{2P}^{\prime}(0)|^2<(2.33\pm1.16)\times10^{-2}$~GeV$^5$, which undershoots
the smallest known potential model result, $0.076$~GeV$^5$
\cite{Eichten:1995ch}, by more than $4\sigma$ and whose central value is more
than a factor of three smaller than the latter.
Conversely, if we choose $|R_{2P}^\prime(0)|^2$ within the ballpark of
potential model calculations, then the upper bound on BR is around three times
smaller than the lower bound 2.6\% \cite{Beringer:1900zz}.
Also the LHCb data point is nicely described by the fit.
Excluding it from the fit appreciably increases the central values and errors
of the fit results and pushes $\chi^2/\mathrm{d.o.f.}$ far below unity, to
0.21.
Specifically, we find
$\langle\mathcal{O}^{\chi_{c1}(2P)}(^3\!P_1^{[1]})\rangle\mathrm{BR}=
(4.94^{+2.39}_{-2.44})\times10^{-3}$~GeV$^5$ and
$\langle\mathcal{O}^{\chi_{c1}(2P)}(^3\!S_1^{[8]})\rangle\mathrm{BR}=
(8.74^{+1.46}_{-1.72})\times10^{-5}$~GeV$^3$.
However, the NLO NRQCD prediction for Eq.~(\ref{LHCb}) now reads 
$(8.04^{+1.42}_{-1.56})$~nb, so that the theory band overshoots the LHCb result
by almost two experimental standard deviations.

We now study the influence of the $\mathcal{O}(v^2)$ corrections on top of the
$\mathcal{O}(\alpha_s)$ ones.
We first observe that, in the kinematic range of our fits, the additional
contributions to Eq.~(\ref{xs}) proportional to
$\langle\mathcal{P}^{\chi_{c1}(2P)}(^3\!S_1^{[8]})\rangle$ and
$\langle\mathcal{P}^{\chi_{c1}(2P)}(^3\!S_1^{[8]},^3\!D_1^{[8]})\rangle$ have
$p_T$ dependencies that match the one of the contribution proportional to
$\langle\mathcal{O}^{\chi_{c1}(2P)}(^3\!S_1^{[8]})\rangle$ within a few
percent, so that these three LDMEs cannot be determined individually.
We thus account for the $\mathcal{O}(v^2)$ corrections by replacing
$\langle\mathcal{O}^{\chi_{c1}(2P)}(^3\!S_1^{[8]})\rangle$ with
$\mathcal{M}_8=\langle\mathcal{O}^{\chi_{c1}(2P)}(^3\!S_1^{[8]})\rangle
+c_1\langle\mathcal{P}^{\chi_{c1}(2P)}(^3\!S_1^{[8]})\rangle/m_c^2
+c_2\langle\mathcal{P}^{\chi_{c1}(2P)}(^3\!S_1^{[8]},^3\!D_1^{[8]})\rangle
/m_c^2$,
with $c_1=-1.06\pm0.03$ and $c_2=0.73\pm0.02$.
As in the one-parameter fit above, we fix
$\langle\mathcal{O}^{\chi_{c1}(2P)}(^3\!P_1^{[1]})\rangle=0.438$~GeV$^5$
\cite{Buchmuller:1980su}.
The fit to the CDF \cite{Acosta:2003zx,Bauer:2004bc}, CMS
\cite{Chatrchyan:2013cld}, and LHCb \cite{Aaij:2011sn} data then yields
$\langle\mathcal{P}^{\chi_{c1}(2P)}(^3\!P_1^{[1]})\rangle/m_c^2
=(0.517\pm0.059)~\mathrm{GeV}^5$ and
$\mathcal{M}_8=(5.71\pm0.32)\times10^{-3}~\mathrm{GeV}^5$ with
$\chi^2/\mathrm{d.o.f.}=2.91/4=0.73$.
I.e.\ the fit is excellent, but the hierarchy
$(\langle\mathcal{P}^{\chi_{c1}(2P)}(^3\!P_1^{[1]})\rangle/m_c^2)/
\langle\mathcal{O}^{\chi_{c1}(2P)}(^3\!P_1^{[1]})\rangle=\mathcal{O}(v^2)$
predicted by the NRQCD velocity scaling rules \cite{Bodwin:1994jh} is strongly
violated.
Including $\langle\mathcal{O}^{\chi_{c1}(2P)}(^3\!P_1^{[1]})\rangle$ among the fit
parameters, we obtain
$\langle\mathcal{O}^{\chi_{c1}(2P)}(^3\!P_1^{[1]})\rangle=(0.432\pm0.286)$~GeV$^5$,
$\langle\mathcal{P}^{\chi_{c1}(2P)}(^3\!P_1^{[1]})\rangle/m_c^2
=(0.509\pm0.438)~\mathrm{GeV}^5$, and
$\mathcal{M}_8=(5.66\pm2.35)\times10^{-3}~\mathrm{GeV}^5$ with
$\chi^2/\mathrm{d.o.f.}=2.91/3=0.97$.
I.e.\ the central values and $\chi^2$ almost go unchanged, while the errors are
magnified.
On the other hand, if we assume that
$\langle\mathcal{P}^{\chi_{c1}(2P)}(^3\!P_1^{[1]})\rangle/m_c^2
=v^2\langle\mathcal{O}^{\chi_{c1}(2P)}(^3\!P_1^{[1]})\rangle$
with $v^2=0.3$, then we obtain
$\langle\mathcal{O}^{\chi_{c1}(2P)}(^3\!P_1^{[1]})\rangle\mathrm{BR}
=(3.39\pm1.25)\times10^{-3}~\mathrm{GeV}^5$
with $\chi^2/\mathrm{d.o.f.}=4.06/4=1.02$.
This corresponds to
$|R_{2P}^{\prime}(0)|^2<(3.03\pm1.12)\times10^{-2}$~GeV$^5$ at 90\% C.L.,
which falls more than $4\sigma$ below the smallest known potential model
result, $0.076$~GeV$^5$ \cite{Eichten:1995ch},

In conclusion, we tested the hypothesis that the $X(3872)$ hadron, whose nature
is remaining undetermined even a decade after its discovery, is a pure
$\chi_{c1}(2P)$ charmonium state, by fitting all available data of prompt
$X(3872)$ production, from the CDF \cite{Acosta:2003zx,Bauer:2004bc},
CMS \cite{Chatrchyan:2013cld}, and LHCb \cite{Aaij:2011sn} Collaborations, at
NLO in $\alpha_s$ and $v^2$ within the effective quantum field theory of NRQCD
endowed with the factorization theorem proposed by Braaten, Bodwin, and Lepage
\cite{Bodwin:1994jh}.
NRQCD factorization, which is arguably the only game in town among the
candidate theories of heavy-quarkonium production and decay, has recently been
impressively consolidated at NLO by global analyses of the world data of
$J/\psi$ inclusive production (for a review, see
Ref.~\cite{Butenschoen:2012qr}).
Assuming the color-singlet LDME
$\langle\mathcal{O}^{\chi_{c1}(2P)}(^3\!P_1^{[1]})\rangle$ to be in the
ballpark of well-established potential models \cite{Eichten:1995ch} and
imposing the lower bound on the BR of $X(3872)\to J/\psi+\pi^+\pi^-$ quoted by
the Particle Data Group \cite{Beringer:1900zz}, we find that the pure
$\chi_{c1}(2P)$ assignment to the  $X(3872)$ hadron is strongly disfavored.
If the $\mathcal{O}(v^2)$ corrections are neglected, the goodness of the fit is
unacceptably poor, and if they are included, the NRQCD velocity scaling rules
\cite{Bodwin:1994jh} are strongly violated.
The tension may be somewhat relaxed by excluding the LHCb data point
\cite{Aaij:2011sn} from the fit, which is, however, unmotivated and
unsatisfactory, the more so as this challenges the CDF
\cite{Acosta:2003zx,Bauer:2004bc} and CMS \cite{Chatrchyan:2013cld}
measurements of the non-prompt $X(3872)$ BR.

If we assume that the $X(3872)$ hadron is a quantum-mechanical superposition of
the $\chi_{c1}(2P)$ meson and a
$D^{\ast0}\overline{D}^0/D^0\overline{D}^{\ast0}$ molecule and that the prompt
production rate of the latter is negligible because of its minuscule binding
energy, then our two-parameter fit including the LHCb data point
\cite{Aaij:2011sn} (see Table~\ref{Table}) allows us to convert the bounds
$|R_{2P}^\prime(0)|^2>0.076$~GeV$^5$ \cite{Eichten:1995ch} and
$\mathrm{BR}>2.6\%$ \cite{Beringer:1900zz} into the bound
$|\langle\chi_{c1}(2P)|X(3872)\rangle|^2<(31\pm15)\%$ on the probability of
encountering the $\chi_{c1}(2P)$ component in the $X(3872)$ state.
If we also include $\mathcal{O}(v^2)$ corrections and enforce the proper
scaling of $\langle\mathcal{P}^{\chi_{c1}(2P)}(^3\!P_1^{[1]})\rangle/m_c^2$
with $v^2$, then we have $|\langle\chi_{c1}(2P)|X(3872)\rangle|^2<(40\pm15)\%$.
Despite concerted experimental and theoretical endeavors during the past
decade, the quest for the ultimate classification of the $X(3872)$ resonance
remains one of the most tantalizing puzzles of hadron spectroscopy at the
present time.

We thank A.~Vairo for a useful communication regarding
Ref.~\cite{Brambilla:2008zg}.
This work was supported in part by
BMBF Grant No.\ 05H12GUE.

{\it Note added.} After submission, a preprint \cite{Meng:2013gga} appeared,
in which $X(3872)$ hadroproduction is studied at NLO in $\alpha_s$ by
performing a two-parameter fit to the CMS data \cite{Chatrchyan:2013cld} and
verifying consistency with the CDF data point \cite{Acosta:2003zx}.
In our notation, the fit results of Ref.~\cite{Meng:2013gga} are 
$\langle\mathcal{O}^{\chi_{c1}(2P)}(^3\!P_{1}^{[1]})\rangle
=(0.17\pm0.07)~\mathrm{GeV}^5$ and
$\langle\mathcal{O}^{\chi_{c1}(2P)}(^3\!S_{1}^{[8]})\rangle
=(3.34\pm1.69)\times10^{-3}~\mathrm{GeV}^3$ with
$\chi^2/\mathrm{d.o.f.}=0.52/2=0.26$, nicely confirming the corresponding
results in the rightmost column of Table~\ref{Table}.

\end{document}